
\input phyzzx

\pubnum={ITP 92-59}
\date={December 10, 1992}
\titlepage

\title{$E_7$ as $D=10$ space-time symmetry --- Origin of the twistor transform}
\author{Martin Cederwall}
\address{Institute of Theoretical Physics \break
Chalmers University of Technology {\rm and} University of
Gothenburg \break S-412 96 Gothenburg, Sweden}
\vfil
\abstract
Massless particle dynamics in $D=10$ Minkowski space is given an
$E_7$-covariant formulation, including both space-time and twistor
variables. $E_7$ contains the conformal algebra as a subalgebra.
Analogous constructions apply to $D=3,4$ and 6.
\submit{Physics Letters B}
\endpage

It is well known that massless particle actions are conformal
invariant (with the possible exception of $D=10$ superparticles, in
which case the matter is not clearly understood). However, there is a
still larger symmetry present, as will be demonstrated in this paper.
Only bosonic particles will be treated. Explicit calculations apply to
$D=10$ -- completely analogous constructions are valid in $D=3,4$ and
6. Some technical details and conventions are found in the appendix.

One has traditionally two options for manifestly conformal
formulations of particle dynamics, the space-time picture and, in
$D=3,4,6$ or 10, the twistor picture. The space-time formulation, on
one hand, with $P_mP^m\approx 0$ as only constraint, is easily made
conformally covariant by enlarging the vectors $X^m,P^m$ of $SO(1,9)$
to become vectors $X^\mu ,P^\mu$ of $SO(2,10)$. The constraints are
taken to be $X_\mu X^\mu\approx 0$, $X_\mu P^\mu\approx 0$ and $P_\mu
P^\mu\approx 0$. With gauge choices $X^\oplus =c$ (=constant) and
$P^\oplus =0$, the ordinary space-time picture is recovered (this of
course applies to any dimensionality).

The twistor picture[1-4], on the other hand, is reached via the
twistor transform
$$\eqalign{P^m=&\,\,\coeff{1}{2}\psi^\alpha\gamma^m_{\alpha\beta}\psi^\beta\cr
\omega_\alpha=&\,\,X_m\gamma^m_{\alpha\beta}\psi^\beta\cr}\eqno(1)$$
which implies that the conformal spinor $Z^A=[\psi^\alpha
,\omega_\alpha]^t$ satisfies seven constraints, generating $S^7$ (a
covariantly conformal form of these constraints is given in ref.4).

Consider now a set of phase-space variables $\Xi^M\in 56$ of $E_7$.
When $E_7\rightarrow Sp(2)\times SO(2,10)$ the the branching rule[5]
is $56\rightarrow (2,12)+(1,32)$ and for the adjoint $133\rightarrow
(3,1)+(2,32')+(1,66)$, so that $\Xi^M=(S^{a\mu},Z^A)$ and $133\ni
T^{\cal A}=(T^{\{ab\}},T^a_{A'},T^{[\mu\nu ]})$. If
$\{\Xi^M,\Xi^N\}=g^{MN}$, $E_7$ is generated by $T^{\cal
A}=\coeff{1}{2}\Xi^M\Omega^{\cal A}_{MN}\Xi^N$, $\Omega^{\cal A}_{MN}$
being Clebsh-Gordan coefficients for $56\times 56\times 133\rightarrow
1$. $S^{a\mu}=(X^\mu ,P^\mu)^t$ is to be interpreted as the conformal
space-time vectors and $Z^A$ as the twistor variables. The set of
constraints is chosen to be $$T^{\cal A}\approx 0\eqno(2)$$ which is
obviously first class. The content of eq.(2) will now be analyzed.

The pure $Sp(2)$ part $T^{ab}\approx 0$ contains exactly the
constraints stated above for the space-time picture. I make the same
gauge choices, with $c=1/\sqrt{2}$ (this reduces manifest covariance
to $SO(1,9)$), solve for $X^\ominus ,P^\ominus$ and insert in the
remaining constraints. The resulting set of equations is highly
reducible. With some help from formulas in the appendix, it is easy to
verify that $T^{\oplus m}\approx 0$ and the second component of
$T^1_{A'}\approx 0$ are exactly the twistor transform equations (1).
Once these are fulfilled, all other identities in eq.(2) hold. The
system described by eq.(2) is thus equivalent to a massless particle
in $D=10$, and may be gauge-fixed to either the space-time or the
twistor picture.

In the dimensionalities $D=3,4$ and 6, the constructions are analogous
to the one above (in $D=4$ and 6 the twistor variables obey bilinear
constraints generating $S^1$ and $S^3$ that also are part of the
symmetry). The resulting algebras are $Sp(6)$, $SU(6)$ and SO(12),
respectively. The algebras can be collectively described as $Sp(6;{\bf
K}_\nu)$ in D=$\nu$+2, where ${\bf K}_\nu$, $\nu$=1,2,4,8 is the
division algebra of dimension $\nu$. They are conformal algebras of
the Jordan algebras of 3$\times$3 hermitean matrices with entries in
${\bf K}_\nu$[6,7]. The conformal algebra in $D=\nu+2$ is $SO(2,\nu
+2)\approx Sp(4;{\bf K}_\nu)$, so one sees how it is contained in the
present larger structure.

The real forms of the algebras are defined by their maximal compact
subalgebras, which are the symmetric subalgebras $SU(3)\times
U(1)\subset Sp(6)$, $SU(3)^2\times U(1)\subset SU(6)$, $SU(6)\times
U(1)\subset SO(12)$ and $E_6\times U(1)\subset E_7$ respectively. It
is worth mentioning that the real form of $SO(12)$ does not belong to
the class $SO(n,12-n)$, and that it, in contrast to the $D=10$
conformal algebra $SO(2,10)$, does posess a superextension[7].

It is probably motivated to regard the extended conformal symmetries
considered in this paper not only as convenient algebraic
constructions, but as fundamental geometric properties of combined
space-time/twistor spaces underlying $D=\nu+2$ Minkowski space.

\appendix
$Sp(2)$ spinor indices are raised and lowered with $\epsilon_{ab}$ as
$y_a=\epsilon_{ab}y^b$. The representation 3 consists of symmetric
matrices $M^{ab}$.

$SO(1,9)$ gamma-matrices are denoted $\gamma^m_{\dot{\alpha}\alpha}$,
$\tilde\gamma^{m\alpha\dot{\alpha}}$ and obey
$$\gamma^{\{m}\tilde\gamma^{n\}}=\eta^{mn}{\bf
1},\;\;\tilde\gamma^{\{m}\gamma^{n\}}=\eta^{mn}{\bf 1}$$ with
$\eta^{mn}=diag(-1,1,\ldots ,1)$.  From these, $SO(2,10)$
gamma-matrices $\Gamma^\mu_{A'A}$ are constructed: $$\Gamma^\oplus
=\left[\matrix{\sqrt{2}\,{\bf 1}&0\cr 0&0\cr}\right],\;\;
\Gamma^\ominus =\left[\matrix{0&0\cr0&\sqrt{2}\,{\bf 1}\cr}\right],\;\;
\Gamma^m =\left[\matrix{0&\tilde\gamma^m\cr\gamma^m&0\cr}\right].$$
Then, $$\Gamma^{\{\mu
A}_{A'}\Gamma^{\nu\}A'}_B=\eta^{\mu\nu}\delta^A_B\;,\;\;
\Gamma^{\{\mu A}_{A'}\Gamma^{\nu\}B'}_A=\eta^{\mu\nu}\delta^{B'}_{A'}$$
with $\eta^{\oplus\ominus}=-1$, $\eta^{mn}$ as above.
$\Gamma^{\mu\nu}_{AB}$ is defined as $\Gamma^{[\mu
A'}_A\Gamma^{\nu]}_{A'B}$. The invariant spinor metrics
$$g_{AB}=\left[\matrix{0&{\bf 1}\cr -{\bf 1}&0\cr}\right]=g_{A'B'}$$
are used to raise and lower spinor indices as $Y_A=g_{AB}Y^B$ and
analogously for primed indices.

The non-zero Clebsh-Gordan coefficients for $56\times 56\rightarrow
133$ of $E_7$ are, with indices according to the maximal subalgebra
$Sp(2)\times SO(2,10)$ (no attention paid to normalization of the
generators): $$\Omega^{ab}_{c\mu
,d\nu}=\coeff{1}{2}\eta_{\mu\nu}\delta^{\{a}_c\delta^{b\}}_d\;,\;\;
\Omega^{aA'}_{b\mu,A}=\Omega^{aA'}_{A,b\mu}=\delta^a_b\Gamma^{A'}_{\mu A}\;,\;
\;
\Omega^{\mu\nu}_{a\kappa,b\lambda}=
\epsilon_{ab}\delta^{\mu\nu}_{\kappa\lambda}\;,\;\;
\Omega^{\mu\nu}_{A,B}=-\coeff{1}{4}\Gamma^{\mu\nu}_{AB}\;.$$
The invariant metric for $56\times 56\rightarrow 1$ is $g_{MN}$, with
$g_{a\mu ,b\nu}=\eta_{\mu\nu}\epsilon_{ab}$ and $g_{AB}$ as above.

\ref{R. Penrose and M.A.H. McCallum
\journal Phys.Rep.&6(72)241 and references therein.}
\ref{I. Bengtsson and M. Cederwall\journal Nucl.Phys.&B302(88)81.}
\ref{N. Berkovits\journal Phys.Lett.&247B(90)45.}
\ref{M. Cederwall\journal J.Math.Phys.&33(92)388.}
\ref{R. Slansky\journal Phys.Rep.&79(81)1.}
\ref{A. Sudbery\journal J.Phys.&A17(84)939.}
\ref{M. Cederwall\journal Phys.Lett&210B(88)169.}

\refout

\end